\documentclass[10pt,journal,final]{IEEEtran}
\IEEEoverridecommandlockouts

\usepackage{amsfonts,amssymb}
\usepackage{array}
\usepackage{ifpdf}
\usepackage{cite}
\usepackage{stfloats}
\usepackage{bm}
\usepackage{verbatim}
\usepackage{color,xcolor}
\usepackage{subfigure}
\usepackage{cite}
\usepackage{subfloat}

\ifCLASSINFOpdf
   \usepackage[pdftex]{graphicx}
   \graphicspath{{../pdf/}{../jpeg/}}
   \DeclareGraphicsExtensions{.pdf,.jpeg,.png}
\else
   \usepackage[dvips]{graphicx}

   \graphicspath{{../eps/}}

   \DeclareGraphicsExtensions{.eps}
\fi

\usepackage[cmex10]{amsmath}

\usepackage{algorithm}
\usepackage{algorithmic}
\usepackage{makecell}

\begin{document}
	%\listoffigures
%	\captionsetup[figure]{labelformat={default},labelsep=period,name={Fig.}}
\title{Achieving Multi-beam Gain in Intelligent Reflecting Surface Assisted Wireless Energy Transfer}
\author{Chi~Qiu, Qingqing~Wu, Meng~Hua, Xinrong~Guan, Yuan~Wu
\thanks{C. Qiu, Q. Wu, M. Hua and Y. Wu are with the State Key Laboratory of Internet of Things for Smart City, University of Macau, Macau, 999078, China (email: yc17434@connect.um.edu.mo; qingqingwu@um.edu.mo; menghua@um.edu.mo; yuanwu@um.edu.mo). X. Guan is with the College of Communications Engineering, Army Engineering University of PLA, Nanjing 210007, China (e-mail: guanxr@aliyun.com).
}}%
\maketitle

\begin{abstract}
Intelligent reflecting surface (IRS) is a promising technology to boost the efficiency of wireless energy transfer (WET) systems. However, for a multiuser WET system, simultaneous multi-beam energy transmission is generally required to achieve the maximum performance, which may not be implemented by using the IRS having only a single set of coefficients. As a result, it remains unknowns how to exploit the IRS to approach such a performance upper bound. To answer this question, we aim to maximize the total harvested energy of a multiuser WET system subject to the user fairness constraints and the non-linear energy harvesting model. We first consider the static IRS beamforming scheme, which shows that the optimal IRS reflection matrix obtained by applying semidefinite relaxation is indeed of high rank in general as the number of energy receivers (ERs) increases, due to which the resulting rank-one solution by applying Gaussian Randomization may lead to significant loss. To achieve the multi-beam gain, we then propose a general time-division based novel framework by exploiting the IRS's dynamic passive beamforming. Moreover, it is able to achieve a good balance between the system performance and complexity by controlling the number of IRS shift patterns. Finally, we also propose a time-division multiple access (TDMA) based passive beamforming design for performance comparison. Simulation results demonstrate the necessity of multi-beam transmission and the superiority of the proposed dynamic IRS beamforming scheme over existing schemes.

\end{abstract}

\section{Introduction}
Intelligent reflecting surface (IRS) has been proposed as a promising solution to reshape the wireless propagation environment by tuning the reflecting elements \cite{1}. This thus motivates recent research on IRS-aided wireless energy transfer (WET) \cite{5} and other related applications \cite{2,4,AIRIS,huameng}. In particular, without the need of any complicated signal processing units, IRS enables intelligent signal reflection over a large aperture to compensate the severe wireless channel attenuation over long distance for improving WET efficiency \cite{11}. \looseness=-1
%\cite{3} @ARTICLE{3,  author={Yu, Guanghua and others},  journal={IEEE Internet of Things J.},   title={Design, Analysis, and Optimization of a Large Intelligent Reflecting Surface-Aided {B5G} Cellular Internet of Things},   year={Sept. 2020},  volume={7},  number={9},  pages={8902-8916},  doi={10.1109/JIOT.2020.2996984}}
\begin{figure}[!t]
	\centerline{\includegraphics[width=3in]{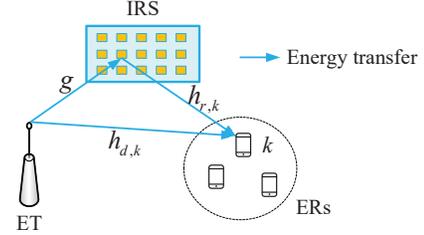}}
	\caption{An IRS-aided WET system.} %--> Fig.1a
	\label{fig1}
	\vspace{-0.3cm}
\end{figure}
\begin{figure}[!t]
	\centerline{\includegraphics[width=3.2in]{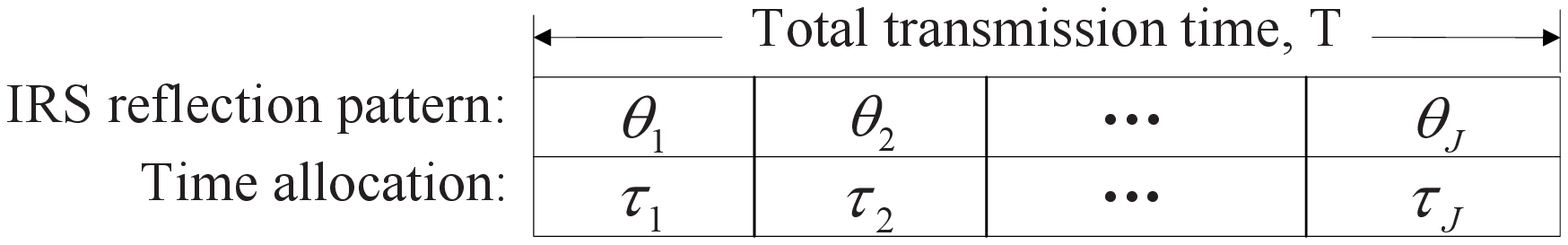}}
	\vspace{-0.2cm}
	\caption{\label{fig2}Transmission protocol of the proposed dynamic IRS beamforming design.}
	\vspace{-0.5cm}
\end{figure}

On the other hand, in multiuser WET system, how to fairly enhance energy transmission efficiency is also an important issue, which has been studied in prior works \cite{xujie,maganggang}. It was shown in \cite{xujie} that the multi-beam energy transmission is the optimal design for the multi-antenna energy transmitter (ET) to maximize the minimum energy harvested by all the energy receivers (ERs). It also reveals that transmitting only one energy beam with adjustable weights over time achieves the same optimal performance as conventional multi-beam design. A similar phenomenon has also been revealed in \cite{maganggang} under non-linear energy harvesting (EH) model. Recall that IRS is only implemented with a single set of coefficients, which, however, is not capable of providing the optimal so-called multi-beam transmission in WET. The above results motivate us to design a series of IRS single beam patterns within the channel coherence time to approach the maximum performance of simultaneous multi-beam transmission, and thus to fully unleash the potential of IRS.

To this end, it still remains an open question that whether there exists performance gain from such time-sharing of IRS reflection patterns and thus needed to be exploited. In this letter, we consider a multiuser IRS-aided  WET system. To avoid severe resource allocation mismatches and significant performance degradation for practical systems, the non-linear EH model is adopted. We first study a special case, where only one reflection pattern is allowed during the whole WET. It is found that the rank of the optimal relaxed IRS reflection coefficient matrix by applying semidefinite relaxation (SDR) indeed grows with the number of ERs in the system, which suggests that a substantial performance will be incurred if a rank-one solution is obtained. To mimic the simultaneous multi-beam transmission, we propose a general framework for dynamic IRS beamforming, where the IRS is allowed to adjust its phase-shift vectors for an arbitrary number of times based on the rank of the reflection matrix from the above to maximize the total harvested energy. Particularly, the advantage of the conventional multiple beams are harnessed via a series of time-varying IRS's reflection patterns. For computational complexity reduction and performance comparison, we then propose a time-division multiple access (TDMA) based beamforming scheme. Specifically, each phase-shift vector is exclusively designed to maximize each ER's harvested energy in each time slot. To solve the formulated direct-current (DC) energy maximization problems for the above cases, we propose efficient algorithms by exploiting successive convex approximation (SCA) technique. Numerical results demonstrate the superiority of the proposed dynamic IRS beamforming scheme, and reveal the upside of time-sharing principle.

%\emph{Notations}: Throughout the paper, boldface upper-case and lower-case letter denote matrix and vector, respectively. $\mathbb{C}^{ a \times b}$ denotes the space of $a\times b$ complex matrices. For complex valued
%vector $\bm x$,  $(\bm x)^H$ and $\operatorname{diag}(\bm x)$ represent Hermitian
%transpose and diagonalization operator of $\bm x$, respectively. For a square matrix $\bm X$, $\operatorname{tr}(\bm X)$ and $\operatorname{rank}(\bm X)$ denote the trace and the rank of $\bm X$, while $\boldsymbol{X} \succeq \bm 0$ indicates that the matrix $\bm X$ is positive semidefinite. For a complex scalar $x$, $|x|$ and $\operatorname{Re}\{x\}$ denote the absolute value and the real part of $x$. $\mathcal{O}(\cdot)$ is the big-O computational complexity notation.

\section{System Model}
As shown in Fig. \ref{fig1}, we consider an IRS assisted WET system, where an IRS with $N$ passive reflecting elements is deployed to assist the WET between a single-antenna ET and a set of single-antenna ERs, denoted by the set $\mathcal{K}=\{1, \dots, K\}$. A quasi-static flat fading channel is considered, and the total available transmission time is given by $T$. We assume that the CSI is perfectly known at the ET by using the state-of-art channel estimation approaches to achieve the system performance upper bound \cite{2}. Denote $h_{d,k}\in  \mathbb{C}^{1\times 1}$, $\mathbf{g}\in \mathbb{C}^{N\times 1}$, and $\mathbf{h}_{r,k}^{H}\in \mathbb{C}^{1\times N}$, $\forall k\in \mathcal{K}$, as the baseband equivalent channels from the ET to ER $k$, from the ET to the IRS, and from the IRS to ER $k$, respectively. As shown in Fig. \ref{fig2}, a novel transmission protocol of the dynamic IRS beamforming design is proposed. To be specific, each channel coherence time $T$ is divided into $J$ time slots, each with duration of $\tau_j$, $j\in{\cal J}=\{1,\dots,J\}$. In each time slot, we assume that the IRS adjusts its reflection pattern individually and ET adjusts its transmit power. To maximize the reflected signal power by the IRS, the reflection amplitude is set to be one. Hence, the reflection coefficient matrix of the IRS in the $j$th time slot is given by $\mathbf{\Phi}_j = \operatorname{diag}\{e^{j\theta_{j,1}},\dots,e^{j\theta_{j,N}}\}\in \mathbb{C}^{N\times N}$, where $\theta_{j,n} \in [0,2\pi), n\in \mathcal{N}=\{1,\dots,N\}$ denotes the phase shift of the $n$th IRS reflecting element.
 
The RF-power received at ER $k$ in time slot $j$ is given by
\vspace{-0.1cm}
\begin{equation}
		P^{r}_{k,j}=P^t_j|\mathbf{h}_{r,k}^H\mathbf{\Phi}_j\mathbf{g}+h_{d,k}^H|^2=P^t_j|\mathbf{q}_{k}^H\bm{\theta}_j+h_{d,k}^H|^2, \label{1}
\end{equation}
where $P^t_j$ represents the allocated transmit power at ET in time slot $j$, $\mathbf{q}_k^H=\mathbf{h}_{r,k}^H \operatorname{diag}(\mathbf{g})$ and $\bm{\theta}_j=[e^{j\theta_{j,1}},\dots,e^{j\theta_{j,N}}]^T$ denotes the IRS passive beamforming vector.
Here, we adopt a practical non-linear EH model to convert the RF power to DC power \cite{9}. As such, the harvested DC power by ER $k$ in the $j$th time slot is given by
\vspace{-0.2cm}
\begin{equation}
	\Phi_{{k,j}}=\frac{X_k}{1+e^{-a_k(P^{r}_{k,j}-b_k)}}-Y_k,
	\vspace{-0.2cm}
\end{equation} where $X_k=M_{k}\left(1+e^{a_{k} b_{k}}\right) / e^{a_{k} b_{k}}$, $Y_k=\frac{M_{k}}{e^{a_{k} b_{k}}}$, $a_k$, $b_k$ and $M_k$, represent the ER $k$'s circuit configurations.

The objective of this letter is to maximize the transferred energy to all ERs subject to energy fairness constraints by jointly optimizing the IRS passive beamforming and resource allocation. We first study \emph{static IRS beamforming} scheme, then propose the \emph{dynamic IRS beamforming} scheme, and finally study the \emph{TDMA-based IRS beamforming} scheme, which are discussed in details as follows.

\subsection{WET with Static IRS Beamforming}
%\begin{figure}[!t]
%	\centerline{\includegraphics[width=2.5in]{rank1}}
%	\caption{\label{rank}Average rank of $\bm\Theta$ versus $K$ with $N=100$ over $100$ channel realizations.} %--> Fig.1a
%	\vspace{-0.5cm}
%\end{figure}
\begin{figure}
	\centering
	\subfigure[Average rank of $\bm\Theta$ versus $K$ with $N=100$.]{\includegraphics[width=2.2in]{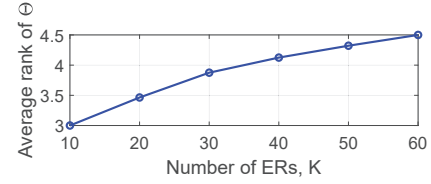}
		\label{rank}}
	\subfigure[The 10 largest average eigenvalues of $\bm\Theta$ with $K=60$.]{\includegraphics[width=2.2in]{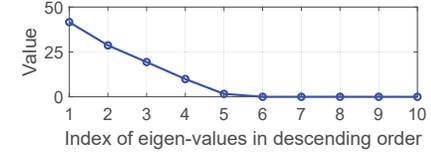}
		\label{eigvalue}}
	\caption{Illustration of properties of $\bm\Theta$.}
	\label{fig3}
	\vspace{-0.4cm}
\end{figure}
%\subsubsection{}
In this case, the phase-shift pattern of the IRS remains unchanged throughout WET. Let $\bm\theta$ denote the phase-shift vector for static IRS beamforming scheme. Thus, the transmit power will also remain constant, denoted by $P$. Thus, the harvested DC power by ER $k$ is $\Phi_{{k}}=\frac{X_k}{1+e^{-a_k(P|\mathbf{q}_{k}^H\bm{\theta}+h_{d,k}^H|^2-b_k)}}-Y_k$. The corresponding optimization problem is formulated as
\vspace{-0.2cm}
\begin{subequations}\label{3}
	\begin{align}
		\max_{P, \bm{\theta}, e} & \ e \\
		\text{ s.t. }
		& T\Phi_{k} \geq \alpha_{k} e, \forall k \in \mathcal{K} \label{3b}, \\
		& PT\leq E_{\operatorname{tot}}, 0\leq P\leq P_{\operatorname{max}} \label{3c},\\
		& |[\bm\theta]_n|=1, \forall n\in \mathcal{N}.\label{3g}
	\end{align}
\end{subequations}
Constraint \eqref{3b} ensures the fairness among all ERs, with $\alpha_k\geq 0$, denoting the target ratio of the $k$th ER's harvested energy to total harvested energy with $ \sum_{k=1}^K \alpha_k=1$. \eqref{3c} denotes the total energy and maximum power constraint, with $E_{\operatorname{tot}}$ denoting the energy budget and $P_{\operatorname{max}}$ representing the maximum allowed transmit power. \eqref{3g} stands for the unit-modulus constraint of the IRS phase-shift.

Note that the above problem is non-convex, we propose an SDR-based algorithm to solve it. Specifically, to fully utilize the available energy at the ET, it can be easily verified that the optimal transmit power is $P= E_{\operatorname{tot}}/{T}$, and thus constraint \eqref{3c} can be dropped. Let $|\mathbf{q}_{k}^H\bm{\theta}_j+h_{d,k}^H|=|\bar{\mathbf{q}}_k^H \bar{\bm{\theta}}_j|$, where $\bar{\bm{\theta}}_j=[\bm{\theta}_j^H \ 1]^H$ and $\bar{\mathbf{q}}_k^H=[\mathbf{q}_k^H \ h_{d,k}^H]$. Then, we can rewrite $P^{r}_{k,j}$ in \eqref{1} as $P^t_j\bar{\bm\theta}_{j}^{H} \mathbf{Q}_{k} \bar{\bm\theta}_{j}$. Define $\mathbf{Q}_k=\bar{\mathbf{q}}_k \bar{\mathbf{q}}^H_k=\left[\begin{array}{ll}
	\mathbf{q}_{k} \mathbf{q}^H_{k} & \mathbf{q}_{k} h_{d,k}^H \\
	h_{d, k} \mathbf{q}_{k}^H & h_{d,k} h_{d,k}^H	\end{array}\right]$ and $\bm{\Theta}=\bar{\bm{\theta}}\bar{\bm{\theta}}^H$, which needs to satisfy $\bm{\Theta}\succeq 0$ and $\operatorname{rank}(\bm\Theta)=1$. Since constraint $\operatorname{rank}(\bm\Theta)=1$ is non-convex, we can apply SDR technique to relax it. As a result, problem \eqref{3} is reduced as
\vspace{-0.2cm}
\begin{subequations}\label{16}
	\begin{align}
		\max_{\bm{\Theta}, e} & \ e \\
		\text{ s.t. }
		& P\operatorname{tr}(\bm{Q}_k\mathbf{\Theta}) \geq b_k-\frac{1}{a_k}\operatorname{ln}\left(\frac{X_k}{\frac{\alpha_ke}{T}+Y_k}\right), \forall k\in \mathcal{K},\\
		& \bm{\Theta}\succeq 0, \\
		& [\bm\Theta]_{n,n}=1, \forall n \in \mathcal{N}.
	\end{align}
\end{subequations}
Although problem \eqref{16} is still non-convex, we observe that with fixed $e$, it is reduced to a linear feasibility check problem. As such, the global optimal solution can be obtained via a bisection/one-dimensional search over $e$. Thus, the solution to problem \eqref{16} serves as an upper bound to problem \eqref{3}. Since the obtained $\bm\Theta$ is not guaranteed to be a rank-one matrix, the Gaussian Randomization (GR) approach can be applied to obtain the sub-optimal solution. As aforementioned, the rank of the IRS reflection coefficient matrix is of interest. For illustration, Fig. \ref{rank} shows the average rank of $\bm\Theta$ versus the number of ERs and Fig. \ref{eigvalue} shows the distribution of the 10 largest average eigenvalues with $K=60$ over $100$ channel realizations, both under the same system setup as will be specified later in Section V. The rank of $\bm\Theta$ is calculated based on the number of eigenvalues exceeds $2\%$ of the maximum one, from which we can see that as $K$ increases, the rank of $\bm\Theta$ also increases, which will result in performance degradation when a rank-one solution is reconstructed by applying GR approach. Motivated by this, we propose \emph{dynamic IRS beamforming} and \emph{TDMA-based IRS beamforming} schemes, elaborated in the next section.
\vspace{-0.2cm}
\section{Dynamic Beamforming Based Multi-beam Design}
In this section, we propose a general time-division based novel framework by exploiting the IRS's dynamic passive beamforming. Then, for performance comparison, we also propose a TDMA-based passive beamforming design.
\vspace{-0.2cm}
\subsection{Dynamic IRS Beamforming}
In this case, the IRS reflection pattern is allowed to adjust for $J$ times, where $J=\operatorname{rank}(\bm\Theta)$ from Section II-A. As such, the IRS generates multi-beams in a time division manner. As introduced early, in this design, we jointly optimize the IRS passive beamforming, time allocation and power allocation. Accordingly, the problem can be formulated as follows
\vspace{-0.2cm}
\begin{subequations}\label{4}
	\begin{align}
		&\max_{\{P^t_j\}, \{\tau_j\}, \{\bm\theta_j\}, e}  e \\
		&\text{ s.t. }
		~\sum_{j=1}^{J} \tau_j \Phi_{{k,j}}\geq \alpha_{k} e, \forall k \in \mathcal{K} \label{4b},\\
		&\qquad \sum_{j=1}^{J} \tau_j \leq T, \tau_{j} \geq 0, \forall j\in \mathcal{J},\label{4c}\\
		&\qquad \sum_{j=1}^{J} P^{t}_{j} \tau_j \leq E_{\operatorname{tot}}, 0\leq P^t_{j}\leq P_{\operatorname{max}}, \forall j\in \mathcal{J}\label{4d},\\
		&\qquad |[\bm\theta_j]_n|=1, \forall n\in \mathcal{N}, \forall j \in \mathcal{J}. \label{4g}
	\end{align}
\end{subequations}
 Constraint \eqref{4c} represents the total and non-zero time constraint.
 \vspace{-0.2cm}
\subsection{TDMA-Based IRS Beamforming}
To reduce computational complexity, the TDMA-based IRS beamforming scheme is proposed. In this design, $T$ is divided into $K$ time slots. In any time slot $k$, the IRS designs its phase-shift pattern to maximize the harvested energy at ER $k$, with $\bm\theta_{k}^*=e^{j(arg(h_{d,k}^H)-arg(\bm q_{k}^H))}$. In this case, only time and power allocations are left to be optimized. Then the harvested DC power by $k$-th ER is give by $\Phi_{{k,j}}^*=\frac{X_k}{1+e^{-a_k(P^t_j|\mathbf{q}_{k}^H\bm{\theta}_j^*\!+\!h_{d,k}^H|^2-b_k)}}-Y_k$. Thus, the problem is reduced to \looseness=-1
\vspace{-0.2cm}
\begin{subequations}\label{6}
	\begin{align}
		&\max_{\{P^t_j\}, \{\tau_j\}, e} e \\
		&\text{ s.t. } ~
		 \sum_{j=1}^{K} \tau_j\Phi_{k,j}^*\geq\alpha_{k} e, \forall k\in \mathcal{K}, \label{6b}\\
		& \qquad \eqref{4c},\eqref{4d}.
	\end{align}
\end{subequations}
The above problem formulations in \eqref{4} and \eqref{6} are both difficult to solve due to the non-convex constraints and highly coupled variables. In the following section, we propose an efficient algorithm, based on SCA techniques to solve them suboptimally.

\section{Proposed solutions}
In this section, efficient algorithms to solve problems \eqref{4} and \eqref{6} are proposed.
\vspace{-0.2cm}
\subsection{Dynamic IRS Beamforming }
Recall that $|\mathbf{q}_{k}^H\bm{\theta}+h_{d,k}^H|^2=\bar{\bm{\theta}}^H \mathbf{Q}_k \bar{\bm{\theta}}$, and by introducing slack variables $z=\{z_{k,j}, j \in \mathcal{J},  k \in \mathcal{K}\}$, constraint \eqref{4b} can be equivalently transformed to
\vspace{-0.2cm}
\begin{equation}\label{8}
	\sum_{j=1}^{J} \tau_j \left(\frac{X_k}{1+z_{k,j}}-Y_k\right)\geq \alpha_{k} e, \forall k\in \mathcal{K},
	\vspace{-0.2cm}
\end{equation}
with additional constraint as
\vspace{-0.1cm}
\begin{equation}\label{7}
	e^{-a_{k}\left(P^t_j\bar{\bm\theta}_{j}^{H} \mathbf{Q}_{k} \bar{\bm\theta}_{j}-b_k\right)} \leq z_{k, j}, \forall j\in \mathcal{J}, \forall k\in \mathcal{K}.
	\vspace{-0.2cm}
\end{equation}
Note that the inequality constraint \eqref{7} is always satisfied with equality for the optimal solution. This can be proved by contradiction. Suppose that the optimal solution is obtained with inequality in \eqref{7}, the value of $z_{k,j}$ can always be decreased to obtain a larger objective value, which is contradictory to the assumption. It can be observed that constraints \eqref{8} and \eqref{7} are still non-convex. However, the left-hand side (LHS) of \eqref{8} is jointly convex with respect to (w.r.t). $\sqrt{\tau_j}$ and $z_{k,j}$. Recall that any convex function is lower-bounded by its first-order Taylor expansion. Thus, the SCA technique is applied. Therefore, for any local points $\tau_j^r$ and $z_{k,j}^r$ in the $r$th iteration, we have
\vspace{-0.2cm}
\begin{align}
		\frac{\left(\sqrt{\tau_{j}}\right)^{2}}{1+z_{k, j}} &\geq \frac{\left(\sqrt{\tau_{j}^{r}}\right)^{2}}{1+z_{k, j}^{r}}-\frac{\left(\sqrt{\tau_{j}^{r}}\right)^{2}}{\left(1+z_{k, j}^{r}\right)^{2}}\left(z_{k, j}-z_{k, j}^{r}\right)\nonumber \\
		&+\frac{2 \sqrt{\tau_{j}^{r}}}{1+z_{k, j}^{r}}\left(\sqrt{\tau_{j}}-\sqrt{\tau_{j}^{r}}\right) \triangleq f_{k, j}^{lb}\left(z_{k, j}, \tau_{j}\right),
\end{align}
which can be readily checked that $f_{k, j}^{lb}$ is jointly convex w.r.t. $z_{k,j}$ and $\tau_j$.
Then, to tackle the non-convexity of \eqref{7}, we first rewrite it into a more simplified form as
\vspace{-0.2cm}
\begin{equation}\label{10}
	P^t_j\bar{\bm\theta}_{j}^{H} \mathbf{Q}_{k} \bar{\bm\theta}_{j}\geq b_k-\frac{\ln z_{k,j}}{a_{k}}, \forall j\in \mathcal{J}, \forall k\in \mathcal{K}.
	\vspace{-0.2cm}
\end{equation}
Note that the optimization variables $P^t_j$ and $\bar{\bm\theta}_j$ are coupled in the LHS of \eqref{10}. Let $P^t_j\bar{\bm{\theta}}_j^H \mathbf{Q}_k \bar{\bm{\theta}}_j=\bar{\bm{\theta}}_j^H \mathbf{Q}_k \bar{\bm{\theta}}_j/ \frac{1}{P^t_j}$. It is observed that although it is not jointly convex w.r.t.  $\bar{\bm{\theta}}_j$ and $P^t_j$, it is jointly convex w.r.t. $\bar{\bm{\theta}}_j$ and $\frac{1}{P^t_j}$, which motivates us to apply SCA to tackle it. Specifically, with any feasible points $\bar{\bm{\theta}}^r_{j}$ and $P^{t,r}_j$, the lower bound of $\bar{\bm{\theta}}_j^H \mathbf{Q}_k \bar{\bm{\theta}}_j/ \frac{1}{P^t_j}$ can be obtained as
\vspace{-0.1cm}
\begin{align}
		&\frac{\bar{\bm{\theta}}_j^H \mathbf{Q}_k \bar{\bm{\theta}}_j}{\frac{1}{P^t_j}} \geq 2 P^{t,r}_j \operatorname{Re}\{\bar{\bm{\theta}}^H_j\mathbf{Q}_k \bar{\bm{\theta}}^r_j\} -P^{t,r}_j \bar{\bm{\theta}}^{rH}_j \mathbf{Q}_k \bar{\bm{\theta}}^r_j\nonumber \\
		& -\bar{\bm{\theta}}^{rH}_j \mathbf{Q}_k \bar{\bm{\theta}}^r_j/\frac{1}{P^{t,r 2}_j}\left(\frac{1}{P^t_j}-\frac{1}{P^{t,r}_j}\right)
		\triangleq g_{k,j}^{lb}\left(\bar{\bm{\theta}}_{j}, P^t_j\right),
\end{align}
which is jointly convex w.r.t. $\bar{\bm{\theta}}_{j}$ and $P^t_j$.

Note that constraint \eqref{4d} is non-convex due to its coupled variables $ P^t_{j}$ and $\tau_j$ in its LHS. We first rewrite $\sum_{j=1}^{J} P^t_{j} \tau_j$ as $\frac{1}{4}\sum_{j=1}^{J}((P^t_j+\tau_j)^2-(P^t_j-\tau_j)^2)$. Since the term $-(P^t_j-\tau_j)^2$ is jointly concave w.r.t. $P^t_j$ and $\tau_j$,  its upper bound can be obtained by applying the SCA technique with given initial points $P^{t,r}_j$ and $\tau^r_j$, yielding
\vspace{-0.1cm}
\begin{align}
		-(P^t_j-\tau_j)^2&\leq -(P^{t,r}_j-\tau^r_j)^2-2(P^{t,r}_j-\tau^r_j)(P^t_j-P^{t,r}_j)\nonumber \\
		&+2(P^{t,r}_j-\tau^r_j)(\tau_j-\tau^r_j)\triangleq \eta^{ub}_{j}\left(P^t_{j}, \tau_{j}\right). \label{11}
\end{align}
Finally, for the non-convex unit-modulus constraint \eqref{4g}, we relax it into a convex form, given by
\vspace{-0.1cm}
\begin{equation}
	|[\bm\theta_j]_n|\leq1, \forall n\in \mathcal{N}, \forall j \in \mathcal{J}.\label{12}
\end{equation}
As a result, given points $z_{k, j}^r$, $\tau_j^r$, $\bar{\bm{\theta}}^r_{j}$ and $P^{t,r}_j$, problem \eqref{4} can be approximated as
\vspace{-0.2cm}
\begin{subequations}\label{13}
	\begin{align}
		&\max_{\{P^t_j\}, \{\tau_j\}, \{\bm\theta_j\}, e} e \\
		&\text{ s.t. } ~
		 \sum_{j=1}^{J}\left( X_k f_{k, j}^{lb}\left(z_{k, j}, \tau_{j}\right)-Y_k\tau_j\right)\geq \alpha_{k} e, \forall k\in \mathcal{K}\label{14b}, \\
		& \qquad g_{k,j}^{lb}\left(\bar{\bm{\theta}}_{j}, P^t_j\right)\geq b_k-\frac{\ln z_{k,j}}{a_{k}}, \forall j\in \mathcal{J}, \forall k\in \mathcal{K},\\
		& \qquad \frac{1}{4}\sum_{j}^{J}((P^t_{j}+\tau_j)^2+\eta^{ub}_{j}\left(P^t_{j}, \tau_{j}\right)) \leq E_{\operatorname{tot}},\label{14d}\\
		& \qquad0\leq P^t_{j}\leq P_{\operatorname{max}}, \forall j\in \mathcal{J},\\
		& \qquad \eqref{4c}, \eqref{12},
		\end{align}
\end{subequations}
which is convex and can be optimally solved by the interior-point method \cite{10}. In addition, since the obtained phase shifts may not satisfy the unit-modulus constraint, the feasible IRS phase-shift vectors are reconstructed by \begin{equation}\label{15}
	[\bm\theta_{j}]_n^*=\frac{[\bm\theta_{j}]_n}{|[\bm\theta_{j}]_n|}, \forall n\in \mathcal{N}, \forall j\in \mathcal{J}.
\end{equation}Finally, time and power allocations are updated by solving \eqref{13} based on the reconstructed phase shifts. It follows that the obtained objective value serves as a lower bound of that of problem \eqref{4}.

%\subsection{Static IRS Beamforming}
%Although the SCA-based algorithm proposed in Section III-A can also be applied to this case, only a lower bound solution can be achieved. To demonstrate the superiority of the proposed dynamic IRS beamforming scheme, an upper bound solution of static IRS scheme is required. In the following, an SDR-based algorithm with 1-D search is proposed. 
%

\subsection{TDMA-Based IRS Beamforming}
Similar to problem \eqref{13}, with the given points $z_{k, j}^r$, $\tau_j^r$ and $P^{t,r}_j$ in the $r$th iteration,  problem \eqref{6} can be approximated as
\vspace{-0.2cm}
\begin{subequations}\label{20}
	\begin{align}
		&\max_{\{P^t_j\}, \{\tau_j\}, e} e \\
		&\text{ s.t. } P^t_j\bar{\bm\theta}_{j}^{*H} \mathbf{Q}_{k} \bar{\bm\theta}_{j}^*\geq b_k-\frac{\ln z_{k,j}}{a_{k}}, \forall j\in \mathcal{J}, \forall k\in \mathcal{K}.\\
		& \qquad 0\leq P^t_{j}\leq P_{\operatorname{max}}, \forall j\in \mathcal{J},\\
		&\qquad \eqref{4c}, \eqref{14b},\eqref{14d},
	\end{align}
\end{subequations}
which can be solved by using the interior-point method \cite{10}. Similarly, the optimal objective value of problem \eqref{20} serves as a lower bound of that of problem \eqref{6}.
\vspace{-0.2cm}
\subsection{Complexity Analysis}
The computational complexity of the two schemes is analyzed as follows. For fairness comparison, the same $J$ is assumed. Specifically, the dynamic IRS beamforming scheme needs to optimize more optimization variables and its complexity is given by $\mathcal{O}\left((2J+NJ)^{3.5}\right)$ based on the analytical results in \cite{12}, whereas the low complexity TDMA-based beamforming scheme mainly needs to solve for problem \eqref{4} with only resource allocation, which results in complexity of $\mathcal{O}\left((2J)^{3.5}\right)$.
%As for the static IRS scheme, though the complexity is given by $\mathcal{O}\left(N^{6.5}\right)$, it reduces the signalling overhead as well as the associated delay for feeding more IRS phase-shift vectors back to the IRS controller.

\section{Numerical Results}
In this section, we provide numerical results to validate the effectiveness of our proposed scheme. In the simulation, a three-dimensional coordinate setup is considered, where the ET is located at $(0,0,0)$ meter (m) and the IRS is located at $(30,0,5)$ m. The ERs are randomly distributed within a circle centered at $(30,0,0)$ m with radius of $5$ m. Rician fading channel is considered for the ET-IRS and IRS-ER links with a Rician factor of $3$ dB, while Rayleigh fading channel is considered for the ET-ER link. The large-scale path loss model is given by $PL(d)=c_{0}\left(d / d_{0}\right)^{-\alpha}$, where $c_{0}=-30$ dB is the path loss at a reference distance of $d_0=1$ m, and path loss exponent $\alpha$ being $2.2$ for the ET-IRS and IRS-ER channels, while $3.6$ for ET-ER channel. We assume all ERs have the same circuit configurations with non-linear EH model parameters given by $a_k=150$, $b_k=0.014$ and $M_k=0.024$ \cite{14}. In addition, the ET antenna gain and ER antenna gain are given by $10$ dBi and $3$ dBi, respectively. Other parameters are set as $N=100$, $\alpha_k=1/K, \forall k, $ $E_{\operatorname{tot}}=10$ Joule (J), $P_{\operatorname{max}}=46$ dBm, $T=1$ s, $d_0=1$ m and $\epsilon=10^{-3}$. We consider the following schemes for performance evaluation: 1) \textbf{Upper bound}: problem \eqref{3} solved by SDR-based algorithm in Section II-A without applying GR; 2) \textbf{Static IRS, GR}: GR is applied to recover the rank-one $\bm\theta$ based on the solution of SDR \cite{2}; 3) \textbf{Static IRS, SCA}: problem \eqref{3} solved by the SCA-based algorithm \cite{huameng}; 4) \textbf{Dynamic IRS}: problem \eqref{4} solved by the SCA-based algorithm proposed in Section IV-A;  5) \textbf{TDMA-based IRS beamforming}: problem \eqref{6} solved in Section IV-B; 6) \textbf{Without IRS}.

\begin{figure}[!t]
	\centerline{\includegraphics[width=2.7in]{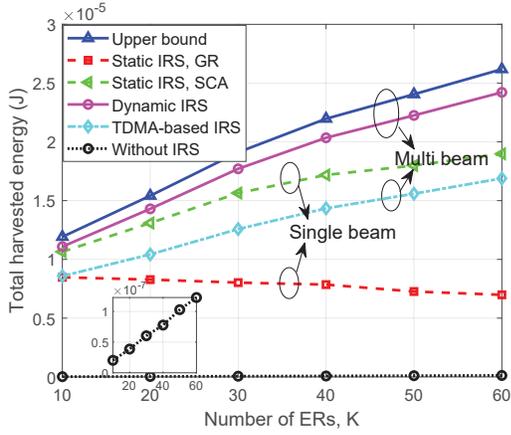}}
	\vspace{-0.2cm}
	\caption{\label{simulation1}Total harvested energy versus $K$.} %--> Fig.1a
	\vspace{-0.5cm}
\end{figure}

Fig. \ref{simulation1} shows the total harvested energy of all ERs versus the number of ERs. It is observed that the IRS-aided WET system outperforms that without IRS, which demonstrates the benefits of integrating IRS into WET system. Particularly, the proposed dynamic IRS beamforming scheme achieves the best performance among all schemes. This is excepted since the IRS is able to achieve the multi-beam gain by proactively generating a series of time-varying beam patterns. It is worth pointing out that the total harvested energy grows as $K$ increases for all schemes, except for the GR approach. This is because that the performance of GR approach significantly degrades from the high rank of matrix $\bm\Theta$ as $K$ increases, which is consistent with analysis in Section II-A. As such, even the proposed TDMA-based scheme with much lower complexity shows superiority over it. Furthermore, though the performance of the SCA-based scheme improves to some extent as compared to the GR-based scheme, it suffers from significant performance loss, which becomes more prominent when $K$ is larger. This is fundamentally because it is essentially still a single-beam based transmission, which thus demonstrates the effectiveness of our proposed dynamic IRS beamforming scheme. It approaches the upper bound with a performance loss of less than 9\%. As compared with other benchmark schemes, for example, with $K=60$, the total harvested energy obtained by dynamic IRS beamforming is about $2.4\times 10^{-5}$ J, which achieves performance gains of 27.5\% higher than the SCA-based static beamforming design, respectively. This thus reveals the advantage of properly designed joint phase-shift patterns optimization and resource allocation, highlighting the necessity of multi-beam transmission.
\begin{figure}[!t]
	\centerline{\includegraphics[width=2.7in]{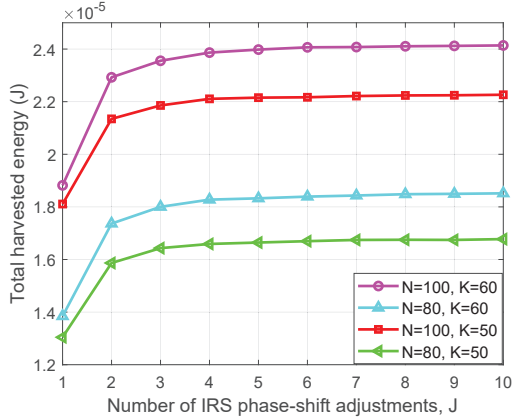}}
	\vspace{-0.2cm}
	\caption{\label{simulation2}Total harvested energy versus $J$ for dynamic IRS beamforming scheme.}
	\vspace{-0.6cm}
\end{figure}

Fig. \ref{simulation2} studies the total harvested energy of all ERs versus the number of IRS reflection patterns for the proposed dynamic IRS beamforming scheme under different system parameters. It is observed that the total harvested energy increases as $J$ increases, and tends to be saturated. This is expected since by dynamically adjusting IRS’s passive beamforming, the system is able to mimic the simultaneous multi-beam transmission through IRS and thus increase the system performance. Furthermore, when the number of IRS reflection patterns, i.e., $J$ exceeds the rank of $\bm\Theta$ (as shown in Fig. \ref{rank}), the performance gain from dynamic IRS beamforming is maximized. Thus the performance improvement becomes negligible when $J$ becomes larger. In addition, as shown in Fig. \ref{eigvalue}, the matrix $\bm\Theta$ is dominated by the first few eigenvalues. As such, even if the average rank of $\bm\Theta=4.5$ with $K=60$, the system readily achieves a good performance with $2$ or $3$ reflection patterns. This implies that in practice, $J$ can be properly chosen to balance between the system performance and the signaling overhead incurred by sending IRS reflection patterns from the ET to IRS.

\section{Conclusion}
In this letter, we investigated the joint passive beamforming and resource allocation optimization for an IRS-aided multiuser WET system under the practical non-linear EH model. We first studied static IRS beamforming scheme, from which we observed that a high rank IRS reflection matrix is generally needed to optimize the multiuser WET with user fairness consideration. To mimic the simultaneous multi-beam transmission with IRS single beam patterns, we then propose the dynamic IRS beamforming scheme and the low-complexity TDMA-based IRS beamforming scheme. The total harvested energy maximization problems were formulated, and efficient algorithms based on SDR and SCA techniques were proposed. Simulation results reveal that the time-division based dynamic IRS beamforming design can improve the WET efficiency and thus provide new insight on the passive beamforming design in multiuser WET systems.
\bibliographystyle{IEEEtran}
\bibliography{reference_short}
\end{document}